\documentstyle[here,epsfig]{article}
\begin{document}
%
%
\newcommand{\lcalcpviol}{
                    \unitlength 0.01cm
                    \begin{picture}(75,75)
		     \put(65,-7){\line(-5,2){38.}}
		     \parbox{3mm}{\mbox{${\cal{L}}_{CP}(x)$}}
		    \end{picture}
                    }
\newcommand{\lcalcp}{\hspace{-0.1cm}
                    \unitlength 0.01cm
                    \begin{picture}(65,30)
		     \put(65,-7){\line(-5,2){38.}}
		     \parbox{3mm}{\mbox{${\cal{L}}_{CP}$}}
		    \end{picture}
                    }
\newcommand{\mcalcpviol}{
                    \unitlength 0.01cm
                    \begin{picture}(90,75)
		     \put(83,-7){\line(-5,2){38.}}
		     \parbox{3mm}{\mbox{${\cal{M}}_{CP}$}}
		    \end{picture}
                    }
\newcommand{\lambdacpviol}{\hspace{-0.1cm}
                    \unitlength 0.01cm
                    \begin{picture}(65,30)
		     \put(65,-7){\line(-5,2){38.}}
		     \parbox{2.5mm}{\mbox{$\Lambda_{CP}$}}
		    \end{picture}
                    }
\newcommand{\lambdacpviolinv}{\hspace{-0.1cm}
                    \unitlength 0.01cm
                    \begin{picture}(65,30)
		     \put(65,-7){\line(-5,2){38.}}
		     \parbox{2.5mm}{\mbox{$\Lambda^{-1}_{CP}$}}
		    \end{picture}
                    }
\newcommand{\lambdacpviolsinv}{\hspace{-0.1cm}
                    \unitlength 0.01cm
                    \begin{picture}(65,30)
		     \put(65,-7){\line(-5,2){38.}}
		     \parbox{2.5mm}{\mbox{$\Lambda^{-2}_{CP}$}}
		    \end{picture}
                    }

\begin{titlepage}

\begin{flushright}
HD-IHEP 96-07   \\
July 1996
\end{flushright}   

\begin{center}
\vspace{1cm}
{\Large\bf Search for \boldmath{$\cal{CP}$}\unboldmath~Violation in the
Decay\\
 \boldmath $ Z \rightarrow b\, {\overline b} \, g$
\unboldmath}\\
{\large \vspace{7ex} Martin Wunsch (University of Heidelberg)\footnote{  
Talk presented at the {\it QCD'96}, Montpellier, France, July 4-12, 1996 }, \\
representing the ALEPH Collaboration }
\end{center}
\vspace{3cm}
\begin{center}
{\bf Abstract}
\end{center}
About three million hadronic decays of the $Z$ collected by ALEPH in the
years 1991 to 1994 are used to search for anomalous $\cal{CP}$
violation beyond the Standard Model in the decay $ Z \rightarrow b 
\bar{b} g$. The study is performed by analyzing angular correlations 
between the two quarks and the gluon in three-jet events and by 
measuring the differential two-jet rate. No signal of $\cal{CP}$~violation 
is found. For the combinations of
anomalous $\cal{CP}$~violating couplings, 
${\hat{h}}_b = {\hat{h}}_{Ab}g_{Vb} -{\hat{h}}_{Vb}g_{Ab}$
and
$h^{\ast}_b=\sqrt{\hat{h}_{Vb}^{2}+\hat{h}_{Ab}^{2}}$
, limits of 
$\mid \hat{h}_b \mid< 0.59$
and
$h^{\ast}_{b} < 3.02$
are given at 95\% CL.
 
\end{titlepage}

\section{Introduction}

The Standard Model predicts only negligible $\cal{CP}$~violating
effects in decays of the $Z$ into quarks or leptons  \cite{Na89}. 
Therefore, a search for $\cal{CP}$~violation at the $Z$
resonance is a test for physics beyond the Standard Model. Simple
extensions \cite{cpbeyond} of the Standard Model with  
$\cal{CP}$-odd and $\cal{CP}$-even Higgs 
states 
are expected to contain $\cal{CP}$~violating 
terms. As couplings 
of Higgs particles are functions of the masses of the interacting 
particles, decays with heavy quarks  deserve 
special interest.
In this analysis \cite{unserpapier} a search for $\cal{CP}$~violating anomalous 
couplings is performed, as proposed by the authors of \cite{Na89,Na94}.  
\section{Theoretical Framework}
In \cite{Na94} $\cal{CP}$-odd couplings are introduced in a model
independent way using an effective Lagrangian. The Standard Model 
Lagrangian density  is extended to include all 
$\cal{CP}$-odd local operators that can be constructed with Standard
Model fields, up to the mass dimension $d=6$.
Effects in $Z\rightarrow b\bar{b}$ from $\cal{CP}$ 
violating  dipole form factors would require the measurement of the spin
directions of the quarks. As for quarks no spin analyzers exist, 
the search for $\cal{CP}$ violation is restricted to the analysis of the   
$\cal{CP}$-odd operator at the vertex
$Z\rightarrow b\bar{b}g$.
In \cite{Na89,Na94} it is shown that all $\cal{CP}$-odd effects are 
proportional to the dimensionless coupling $\hat{h}_b$:
\begin{equation}
\label{hbshape}
{\hat{h}}_b = {\hat{h}}_{Ab}g_{Vb} -
{\hat{h}}_{Vb}g_{Ab}
\end{equation}
with
\[\centering
\hat{h}_{Ab/Vb} = h_{Ab/Vb} \frac{\sin\vartheta_W 
\cos\vartheta_W m^2_Z}{eg_s} \: ,
\]
where  
 $g_s$ is the strong coupling and 
$g_{Vb}$, $g_{Ab}$ are the Standard Model vector and axial vector
couplings of the $b$~quark to the $Z$.\\ 
 
\par
In a first study, $\cal{CP}$-odd variables are analyzed, probing the 
coupling ${\hat{h}}_b$. To measure $ {\hat{h}}_b$ the 
$\cal{CP}$-odd tensor $T'_{ij}$ \cite{Na94} is used:
\begin{equation}\label{eqchap2tij}
T'_{ij}=(\hat{k}_{\bar{q}}-\hat{k}_{q})_i
\left(\frac{\hat{k}_{\bar{q}}\times\hat{k}_{q}}
{\mid\hat{k}_{\bar{q}}\times\hat{k}_{q}\mid}\right)_j + 
(i \leftrightarrow j) \: , 
\end{equation}
where $\hat{k}_{q}$, $\hat{k}_{\bar{q}}$ denote the normalized momentum
vectors of the quark and anti-quark, respectively, 
and $i$, $j$ are the cartesian 
coordinates with $i$, $j$ $=3$ defined to be along the beam axis. 
 The tensor $T'_{ij}$ is symmetric in $i$ and $j$, traceless, and
invariant when exchanging $\hat{k}_{q}$ and $\hat{k}_{\bar{q}}$.
Therefore only the gluon jet has to be tagged. 
If $\cal{CP}$ is violated then $\langle T'_{ij} \rangle \not= 0$.
The most sensitive observable is
$\langle T'_{33} \rangle$  \cite{unserpapier}, it is related to  
$\hat h_b$ \cite{Na94} by:
\begin{equation}
\label{sens} 
\langle T'_{33} \rangle= {\hat{h}}_b {Y'} \: .
\end{equation}
The sensitivity $Y'$ is a constant derived by integration of
the experimentally accessible
phase space.  
 
\par
In a subsequent analysis, additional contributions from anomalous couplings
to the three-jet rate in
the process $Z \rightarrow b\bar{b}g$  have been
searched for. Such additional contributions are proportional to the 
combination ${h^{\ast}_b}^2$, with
\begin{equation}
\label{hsshape}
h^{\ast}_b=\sqrt{\hat{h}_{Vb}^{2}+\hat{h}_{Ab}^{2}}\: ,
\end{equation}
and would manifest themselves in a higher value 
of the strong coupling constant for $b$~quarks, 
$\alpha_{s}^{b}(M_{Z}^{2})$ (for a complementary but related study see 
\cite{afs}).
The analysis of $h^{\ast}_{b}$ is based on the differential two-jet rate
$D_{2}$, which is the normalized distribution of the event shape
variable $Y_{3}$, $D_{2}(Y_{3})=1/\sigma \; d\sigma(Y_{3})/dY_{3}$,
with $Y_{3}$ being the $y_{\textrm{cut}}$ at which an event changes its classification
from three jets to two jets.  
Additional couplings at the vertex $Z \rightarrow b\bar{b}g$ could 
also contribute to the
partial width $\Gamma_{b\bar b}$, thus possibly enhancing $R_b$ beyond
its Standard Model value \cite{Na94}. By attributing the observed deviation of $R^{exp}_b =0.2209 \pm 0.0021$
 from the Standard Model expectation $R_b =0.2155 \pm 0.0005$
\cite{Rb_th} entirely to
these anomalous couplings, a value of $h^{\ast}_b = 1.93 \pm 0.75$ can
be calculated. 
\section{Data Reduction}
The ALEPH detector and its performance are described in detail in  
 \cite{ALEPHdet}. The selection of hadronic
events is detailed in \cite{unserpapier}.
Jets are defined using the JADE or Durham 
algorithms \cite{JADE}. In
the $\hat{h}_b$ analysis the JADE 
algorithm with a fixed cut-off value of $y_{\textrm{cut}}=0.03$ is used. In the
$h^{\ast}_b$ analysis both jet schemes are employed to measure
the differential two-jet rate $D_2(Y_3)$.
The $b$~events are selected exploiting the 
sizeable impact parameters due to $b$~decays.
The $b$~tagging algorithm is described in detail in 
reference \cite{ALEPHbtag}. The probability for each charged particle to
originate from the primary vertex is calculated using the track impact
parameter. They are combined into
probabilities $\cal{P}$ associated to sets of tracks like jets ($\cal{P}_J$)
or whole events ($\cal{P}_E$). In both analyses $b$~events
are selected with a cut on the event probability $\cal{P}_E$, 
resulting in $b$~purities of more than 85\%. 
\section{Analysis of Anomalous Couplings from {\boldmath $Z \rightarrow b\bar{b}g$ \unboldmath} Topologies}
In the study of $\cal{CP}$-odd contributions three-jet events are selected as
described in \cite{unserpapier}. 
The energies of the three reconstructed jets are ordered,
$E_{\textrm{jet}\,1}>E_{\textrm{jet}\,2}>E_{\textrm{jet}\,3}$. 
 In the $b$~sample the gluon jet candidate
is selected using the probability $\cal{P}_J$ of the individual jets:
from the two lower energy jets, the jet with the higher $\cal{P}_J$ 
is chosen as the gluon jet candidate.  
\par
The purity of the sample is estimated using simulated
events generated with JETSET~\cite{Jetset}.
The matching procedure of partons to jets found after reconstruction is detailed in
\cite{unserpapier}.
A purity of   $73.6\% \pm 0.4_{stat} \pm 0.9_{syst}$
 is achieved. The efficiency,
defined as selected $b \bar bg$ events with successfully tagged gluon
compared to the total rate of $Z \rightarrow b \bar bg$ events, is
about 19\%. 
The measurement is performed on the data collected until
1994. After applying the data selection, a sample
of 85342 $b \bar bg$ candidates remains for the $T'_{33}$ measurement.
The following mean value   of $\langle T'_{33}\rangle$ is obtained:
\begin{displaymath}
\langle T'_{33} \rangle = (-0.5 \pm 3.7_{\textrm{stat}}) \times 10^{-3} 
\end{displaymath}

\par
The measurement is also performed on a sample enriched with
light quarks to ensure that no effect due to the selection 
mechanism is present and to check the assumption that no 
$\cal{CP}$~violating effect exists in events with light quarks.
Light flavours are selected with the cut  $\cal{P}_E > 0.5 $.
The gluon jet in this sample is defined to be the one having the lowest energy
in the event. 
The following mean value of $\langle T'_{33}\rangle$ is obtained:
\begin{displaymath}
\langle T'_{33} \rangle = (-0.9 \pm 3.1_{\textrm{stat}} ) \times 10^{-3} 
\end{displaymath}
Obviously  no 
significant discrepancy between the two completely
independent samples is found. Hence it is concluded that no 
significant fake $\cal{CP}$~violating effect due to the selection
is present.\\ 

\subsection{Systematic Errors}
 
To estimate the effect of the selection cuts, all  cuts are varied
in a wide range. Only  those applied
to select well-defined three-jet configurations  ($y_{\textrm{cut}}$, aplanarity, 
jet energies, multiplicities and angles)
have a sizeable influence on the selected sample. Therefore, the systematic 
error is estimated using these cuts only \cite{unserpapier}.
The $\cal{CP}$~invariance of the $b\,\textrm{tag}$ is checked by
dividing the three-jet sample into  disjoint subsamples with different 
$\cal{P}_E$~values. Using these subsamples, a sample of 
$\langle T'_{33}\rangle $~values is measured to test the bias 
of the $\cal{P}_{\textrm{E}}$~cut. The ${\langle T'_{33} \rangle}$~values are 
found to 
be well compatible with
statistical fluctuations. The uncertainty is estimated from a 
Gaussian fit to the distribution.
The $\cal{CP}$~symmetry of the detector is estimated by measuring 
$\langle T'_{33} \rangle$ on a sample of track pairs
in hadronic events \cite{unserpapier}.   The resulting 
asymmetry is compatible with zero within one standard deviation. 
The quality of the reconstruction of the jet direction is estimated by
measuring the difference of the $\langle T'_{33} \rangle$
at the parton level 
and after the full detector simulation. 
The error is derived from the width of the difference distribution. 
A more detailed description of the systematic uncertainties is given in
\cite{unserpapier}.
\subsection{Determination of \boldmath $\hat{h}_b$ \unboldmath}
To extract the coupling $\hat{h}_b$ from the measurement of 
$\langle T'_{33} \rangle$ the effective sensitivity $Y'$ has to be
calculated. This has been done by means of  a Monte Carlo
generator \cite{BernMC}, which includes the $\cal{CP}$~violating
couplings.   The  calculated
sensitivity $Y' = -0.0167 \pm 0.0002_{\textrm{stat}}$ 
is constant, showing no dependence on $\hat{h}_b$ \cite{unserpapier}.
The systematic errors on the sensitivity stem from the errors
on the tagging purity and from the influence of the $b$~quark mass. 
The largest error is caused by
the dependence of the sensitivity on the $b$~quark mass \cite{haberl}. 
Taking into account the total systematic uncertainty \cite{unserpapier} 
the sensitivity is given by 
\[
\centering
Y'  =-0.0167 \pm 0.0002_{\textrm{stat}} \pm 0.0015_{\textrm{syst}}\:.
\]
 \subsection{Results of the \boldmath $\hat{h}_b $ 
\unboldmath Analysis}
Taking the systematic errors into account the measurement of the
$\cal{CP}$-odd observable in a sample of 
$Z \rightarrow b \bar{b} g$ events yields: 
\begin{displaymath}
\langle T'_{33} \rangle = (-0.5 \pm 3.7_{\textrm{stat}} \pm
3.3_{\textrm{syst}}) \times 10^{-3}\:.
\end{displaymath}
The measurement is consistent with $\langle T'_{33} \rangle = 0$.
 The size of the 
$\cal{CP}$-odd coupling $\hat{h}_b$ is extracted using 
eq.~\ref{sens}: 
\begin{displaymath}
\hat{h}_b = 0.03 \pm 0.22_{\textrm{stat}} \pm 0.20_{\textrm{syst}}\: . 
\end{displaymath}
From this
measurement a limit on the coupling of $\mid \hat{h}_b \mid< 0.59$ (95\%
CL) is derived.
\section{Measurement of Anomalous Couplings from the Differential Two-Jet
Rate}
Of all the event shape variables studied in \cite{afs}, 
only for $Y_{3}$ theoretical calculations including anomalous couplings are
available \cite{haberl}. Therefore,
the second analysis concentrates on possible additional contributions
to the differential two-jet rate $D_{2}(Y_{3})$ due to the anomalous coupling 
$h^{\ast}_{b}$. The theoretical prediction for $D_{2}(Y_{3})$ is given by
\begin{eqnarray*}\label{hasko2}
D_{2}^{b}(Y_{3}) = 
\frac{\alpha_{s}(\mu^{2})}{2\pi}\Bigg( A^{\prime}(Y_{3})
+{h^{\ast}_{b}}^{2}C^{\prime}(Y_{3})\Bigg) \nonumber \\ 
+\left(\frac{\alpha_{s}(\mu^{2})}{2\pi}\right)^{2}
\left[A^{\prime}(Y_{3})2\pi
b_{0}\ln\left(\frac{\mu^{2}}{M_{Z}^{2}}\right)+B^{\prime}(Y_{3})\right]
\end{eqnarray*}
where $b_{0}=(33-2n_{f})/12\pi$, $\mu$ is the renormalization scale, and
the coefficients$A^{\prime}$ and $B^{\prime}$ 
 have been computed to second order of perturbative
QCD \cite{AandB}. Here $h^{\ast}_b \gg h^{\ast}_{udsc}$ 
is assumed. Hence the prediction for $udsc$\,quarks  
is entirely fixed by the Standard Model and does not depend on 
$h^{\ast}$. In contrast, $D_{2}$
for $b$~quarks receives contributions from new physics and the prediction
is modified by an additional term proportional to ${h^{\ast}_{b}}^2$.
Note that in the coefficients $A^{\prime}$ and $B^{\prime}$    the
total cross section needed for normalization is changed if anomalous
couplings are present.  
The coefficient $C^{\prime}$ has been calculated
to leading order \cite{haberl}.
\par
A theoretical description
$R^{\textrm{th}}$ of the observable can be derived:
\begin{eqnarray*}\label{hasko3} R^{th} =
\frac{P_{b}D_{2}^{b,b\,\textrm{tag}}(Y_{3})+(1-P_{b})
D_{2}^{udsc,b\,\textrm{tag}}(Y_{3})}
{R_{b}D_{2}^{b\,\textrm{incl}}(Y_{3})+(1-R_{b})D_{2}^{udsc\,\textrm{incl}}(Y_{3})}  
\end{eqnarray*}
where $P_{b}$ is the purity of the lifetime-enriched
 $b$~sample, $R_{b}$ is the fraction of $Z$'s decaying into $b$~quarks and
 $D_{2}^{q\,{\it tag}}$ stands for the distribution of a 
flavour $q$ in a sample of type {\it tag}.  These distributions are constructed from the parton
level predictions and have to be corrected for the following 
effects: mass corrections for $b$~quarks, initial and final state radiation, hadronization effects,
the detector acceptance, the influence of the detector resolution, and
the tagging bias. After applying these corrections to 
$R^{\textrm{th}}$, a value of $h^{\ast}_{b}$
is extracted from a binned least-square fit to the data.
\subsection{ {\boldmath $h^{\ast}_{b}$ \unboldmath} and Systematic Errors}
The fit function $R^{\textrm{th}}$ depends explicitly on
$h^{\ast}_{b}$, $\alpha_{s}$ and $\mu$. 
The value
of the strong coupling constant is set in the fit function to 
 $\alpha_{s}(M^2_Z) = 0.118 $ 
and is varied by  $\pm0.007$.
The renormalization scale is set to 15\,GeV.
This symmetrizes the error from the scale uncertainty, which is
varied from $\mu=m_{b}$ to $\mu=M_{Z}$. 
The fit range  is chosen to optimize the 
sensitivity and to guarantee a good perturbative description. To take into account 
systematic uncertainties due to the limited fit range
the fit is repeated with
a range modified by two bins. 
Jets are reconstructed using charged tracks and neutral calorimeter 
objects.  The uncertainty of the reconstruction procedure is estimated
 by repeating the analysis with charged tracks only. 
The $b$~sample is enriched by means of a  lifetime
tag, which leads to a distortion of the differential two-jet rate
of less than 10\%. Therefore, a correction is elaborated using
full detector simulation.  The stability of these corrections is
checked by varying the lifetime cuts in the data resulting in a change
of the purity of the sample of 10\%.  The same cuts are applied
to the simulated data and the corrections for the tagging bias
recalculated. 
The $b$~quark fragmentation is described
by the fragmentation function of Peterson et al.~\cite{Peterson}.
The main parameter of this function is $\epsilon_{b}$, measured 
to be $\epsilon_{b}=(3.2 \pm 1.7)\times10^{-3}$ \cite{Eps_b}.
Monte Carlo simulations with a corresponding range of values are done to
study the effect on $h^{\ast}_{b}$. 
Another
source of uncertainty is related to mass corrections.  These have been
calculated in \cite{Masef} at tree level.  These calculations are
only complete to $\cal O(\alpha_{s})$ and are applied to
the coefficient $A^{\prime}$.  The uncertainty on
the $b$~quark mass is set to 0.5 $\textrm{GeV}/c^2$ and the corresponding
correction recalculated. 
 In order to account for missing higher
orders, the available $\cal O(\alpha_{s}^{2})$ four-jet computation is
used for correction as well and the difference to the $\cal
O(\alpha_{s})$ result is taken as systematic error. 
Finally, the
parameters $R_{b}$ and $P_{b}$ are varied in the fit function 
within their errors and an error on the normalization is
derived. The typical systematic uncertainty of the cluster algorithms is 
$\Delta  h^{\ast}_{b}=0.38$ \cite{unserpapier}. 
 
\subsection{Hadronization Model Uncertainty and Results of the {\boldmath $h^{\ast}_{b}$\unboldmath} Analysis} 
The main theoretical uncertainty is caused by the hadronization.
 The results achieved with
four different generators are compared in table~\ref{Models}:  the 
matrix element with string fragmentation (ME), 
the parton shower ($Q_{0} = 1$ GeV, being the
cut-off of the parton shower) as  implemented in
JETSET \cite{Jetset} with string fragmentation (PS),  the model of
cluster fragmentation in HERWIG \cite{Herwig} (HW)
and  the dipole cascade model implemented in ARIADNE \cite{Ariadne} (AR). 
 Taking into account the correlations between the two algorithms,
 a combined result for each hadronization model
can be derived which minimizes the
total statistical error.  
\begin{table}[htb]
\begin{center}
\begin{tabular}{c c }\hline 
Model &  $h^{\ast}_{b}$  \\ \hline  
 ME &$1.05 \pm 0.26 \pm 0.40 $\\  
AR & $0.79 \pm 0.30 \pm 0.40 $\\  
PS & $1.75 \pm 0.16 \pm 0.35 $ \\  
HW & $1.79 \pm 0.16 \pm 0.36 $ \\ \hline 
\end{tabular}
\caption[Results for four different models]
{\label{Models} Results on $h^{\ast}_{b}$ from the 
least-square fit for four different hadronization models. The first error is the statistical one, 
the second the total systematic
uncertainty.}
\end{center}
\end{table}
\par
After averaging the combined results of the different hadronization models 
the following result ist obtained:
\begin{displaymath}
h^{\ast}_b = 1.34 \pm 0.22_{\textrm{stat}} \pm 0.38_{\textrm{syst}} \pm 0.50_{\textrm{hadr}}\: . 
\end{displaymath}
where the first error is the statistical one, the second error contains all
systematic uncertainties but 
the hadronization model and the last one reflects the systematic uncertainty
due to  the hadronization model itself. The upper limit is derived by 
adding linearly the systematic and 
statistical errors: 
\begin{displaymath}
h^{\ast}_{b} < 3.02 \mbox{ (95\% CL) } .
\end{displaymath}

\section{Conclusions}
A search for $\cal{CP}$~violation beyond the standard model in the decay 
$Z \rightarrow b \bar{b}g$ has been performed. Two combinations
of $\cal{CP}$-odd couplings, namely $h^{\ast}_b$ and
$\hat{h}_b$ have been analyzed.  
No evidence for $\cal{CP}$-odd couplings is found in both
analyses. The derived limit on $h^{\ast}_b$ is consistent 
with the value calculated from the $R_b$ measurement.
\par
\begin{figure}[H]
\begin{center}
\mbox{\epsfig{file=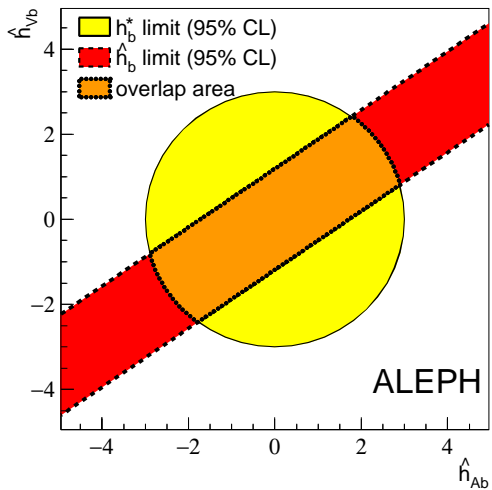,height=6.5cm }}
\caption{ Combined results. The shaded areas depict the constraints 
of the measurements on the couplings $\hat{h}_{Ab}$ and $\hat{h}_{Vb}$.}
\label{combined}
\end{center}
\end{figure}

Using eq.~\ref{hbshape} and eq.~\ref{hsshape}, the
two measurements presented in this paper can be used to constrain
the couplings $\hat{h}_{Ab}$, $\hat{h}_{Vb}$. In
fig.~\ref{combined} the 95\% CL limits of both measurements on $\hat{h}_{Ab}$ and
$\hat{h}_{Vb}$ 
are shown, being well consistent with each other. 

\section*{Acknowledgments}
I want to thank the organizers of the QCD 96 for having set up and conducted 
a very valuable meeting.
I would like to acknowledge the continuous interest and the helpful support
of W. Bernreuther, P. Haberl and O. Nachtmann in this work. 


\begin{thebibliography}{99}
%
%
\bibitem{Na89} W.~Bernreuther, U.~L\"ow, J.~P.~Ma, and O.~Nachtmann, Z.~Phys.~
{\bf C43} (89) 117.
%
%
%
\bibitem{cpbeyond} W. Bernreuther, T. Schr\"oder, and T. N. Pham, Phys. Lett. {\bf B279}
                (92) 389.;
              S. M. Barr, A. Masiero, Phys. Rev. Lett. {\bf 187} (87) 187;
              J. C. Pati, A. Salam, Phys. Rev. {\bf D10} (75) 275.
\bibitem{unserpapier} D.~Buskulic et al., ALEPH Collaboration, CERN/PPE 96-071 (96), 
subm. to Phys.~Lett.~B.
%
%
%
\bibitem{Na94} W.~Bernreuther, D.~Bru\ss, P.~Haberl, and O.~Nachtmann, Z.~Phys.~
{\bf C68} (95) 73.
%
%
%
 
\bibitem{afs} D.~Buskulic et al., ALEPH Collaboration, Phys.~Lett.~{\bf B355} (95) 381;
              H.~Stenzel, PhD thesis, Heidelberg, HD-IHEP 95-03 (95) and references therein
               
%
%
%
\bibitem{Rb_th} 
The LEP Collaborations ALEPH, DELPHI, L3 and OPAL, CERN-PPE/96-017 (96), 
subm. to Nucl. Instr. and Meth.; CERN-PPE/95-172 (95). 
%
%
\bibitem{ALEPHdet} D.~Decamp et al., ALEPH Collaboration, Nucl.~Instr.~and Meth.~
{\bf A294} (90) 121; Nucl.~Instr.~and Meth.~
{\bf A360} (95) 481.
%
%
%
%
%
\bibitem{JADE} W.~Bartel et al., JADE Collaboration, Z.~Phys.~{\bf C33} (86) 23; Phys.~Lett.~{\bf B213} (88) 235;
 S.~Catani et al., Phys.~Lett.~{\bf B269} (91) 432; W.~J.~Stirling, J.~Phys.~{\bf G17} (91) 1657.
%
%
%
\bibitem{ALEPHbtag} D.~Buskulic et al., ALEPH Collaboration, Phys.~Lett.~{\bf B313}
    (93) 535.
%
%
%
\bibitem{Jetset} T.~Sj\"ostrand, Comp.~Phys.~Comm.~{\bf 39} (86) 347; {\bf 43} (87) 367; {\bf 82} (94) 74.
%
%
%
\bibitem{BernMC} W.~Bernreuther,  private communication
%
%
%
\bibitem{haberl} P.~Haberl, HD-THEP 96-16 (96) 
%
%
%
\bibitem{AandB} Z.~Kunszt et al., in ``Z Physics at LEP 1'', edited by G.~Altarelli, CERN Yellow Report 89-08 (89) 373.
%
%
%
\bibitem{Peterson} C.~Peterson et al., Phys.~Rev.~{\bf D27} (83) 105.
%
%
%
\bibitem{Eps_b}  D.~Buskulic et al., ALEPH Collaboration, Z.~Phys.~{\bf C62} (94) 179.
%
%
%
\bibitem{Masef} A.~Ballestrero et al., Phys.~Lett.~{\bf B294} (92) 425; {\bf B323} (94) 53;  {\bf B415} (94) 265.
%
%
%
\bibitem{Herwig} G.~Marchesini et al., Nucl.~Phys.~{\bf B310} (88) 571; Comp.~Phys.~Comm.~{\bf 67} (92) 269.
%
%
%
\bibitem{Ariadne} L.~L\"onnblad, Comp.~Phys.~Comm.~{\bf 71} (92) 465.
%
%
%
\end{thebibliography}
\end{document}